\documentclass[a4paper]{jpconf}
\usepackage{amsmath,amssymb}
\usepackage{algorithm,algpseudocode}
\usepackage{graphicx,wrapfig,color}
\usepackage{multicol}
\makeatletter
\newenvironment{figurehere}
{\def\@captype{figure}}
{}
\makeatother
\begin{document}
\title{Multicanonical simulation of the Domb-Joyce model and the G\={o} model: new enumeration methods for self-avoiding walks}

\author{Nobu C. Shirai$^{1,2}$ and Macoto Kikuchi$^{2,1,3}$}

\address{$^{1}$ Graduate School of Science, Osaka University, Toyonaka, Osaka 560-0043, Japan}
\address{$^{2}$ Cybermedia Center, Osaka University, Toyonaka, Osaka 560-0043, Japan}
\address{$^{3}$ Graduate School of Frontier Biosciences, Osaka University, Suita, Osaka 565-0871, Japan}
\ead{shirai@cp.cmc.osaka-u.ac.jp}

\begin{abstract}
We develop statistical enumeration methods for self-avoiding walks using a powerful sampling technique called the multicanonical Monte Carlo method.
Using these methods, we estimate the numbers of the two dimensional $N$-step self-avoiding walks up to $N=256$ with statistical errors.
The developed methods are based on statistical mechanical models of paths which include self-avoiding walks.
The criterion for selecting a suitable model for enumerating self-avoiding walks is whether or not the configuration space of the model includes a set for which the number of the elements can be exactly counted.
We call this set a scale fixing set.
We selected the following two models which satisfy the criterion: the G\={o} model for lattice proteins and the Domb-Joyce model for generalized random walks. There is a contrast between these two models in the structures of the configuration space.
The configuration space of the G\={o} model is defined as the universal set of self-avoiding walks, and the set of the ground state conformation provides a scale fixing set.
On the other hand, the configuration space of the Domb-Joyce model is defined as the universal set of random walks which can be used as a scale fixing set, and the set of the ground state conformation is the same as the universal set of self-avoiding walks.
From the perspective of enumeration performance, we conclude that the Domb-Joyce model is the better of the two. The reason for the performance difference is partly explained by the existence of the first-order phase transition of the G\={o} model.
\end{abstract}

\section{Introduction\label{intro}}
A self-avoiding walk (SAW) is a path on a lattice with the spatial restriction that the path must not visit the same site more than once~\cite{MadrasSlade1993}. 
A SAW is used as a model of chain polymers, because the above restriction represents the excluded volume effect.
Lattice protein models such as the G\={o} model~\cite{TaketomiUedaGo1975,GoTaketomi1978,Go1983} and the HP model~\cite{LauDill1989} have played important roles in the construction of the funnel picture of the energy landscape~\cite{Go1983,OnuchicWolynes1997,DillChan1997}, which forms the basic theoretical understanding of protein folding.

Many sophisticated methods have been developed to analyze the thermodynamic properties of a lattice protein.
The multi-self-overlap ensemble (MSOE)~\cite{IbaKikuchi1998,ChikenjiIba1999}, which is an extended version of the multicanonical Monte Carlo method~\cite{BergNeuhaus1991,BergNeuhaus1992}, is one of the best methods to find a ground state and calculate thermodynamic properties of a long lattice protein. The main idea of the MSOE is that the self-avoiding condition is relaxed so that some intersections of a chain are allowed. In the present work, we apply the idea to the enumeration problem of finite-step SAWs.

Besides having many applications like polymers, the SAW itself has many interesting asymptotic behaviors of infinite steps and has been studied by physicists, chemists, and mathematicians for a long time. In spite of great efforts, major parts of proposed asymptotic behaviors have not been solved in the rigorous mathematical sense and remain as conjectures. Enumeration of $N$-step SAWs is a famous unsolved problem and the exact number of two-dimensional SAWs is known up to only 71 steps~\cite{Jensen2004}. The total number of $N$-step SAWs is written as $c_N$, and $c_{71}$ is counted as
\[
c_{71}= 4\ 190\ 893\ 020\ 903\ 935\ 054\ 619\ 120\ 005\ 916
\simeq 4.1909\times 10^{30},
\]
which is larger than Avogadro's constant ($\simeq 6.0221\times 10^{23}$).
Since $c_N$ increases exponentially with $N$, it will soon be impossible to deal with all components of $N$-step SAWs on a computer and we have to use statistical methods for approximate enumeration of $c_N$ with large $N$.

In our study, in order to estimate $c_N$ with large $N$ accurately, we developed new enumeration methods using multicanonical simulations of the following two kinds of statistical mechanical models: the Domb-Joyce model~\cite{DombJoyce1972} for generalized random walks and the G\={o} model~\cite{TaketomiUedaGo1975,GoTaketomi1978,Go1983} for lattice proteins.
By using our methods, we were able to estimate the number of SAWs on a square lattice up to 256 steps with error estimation.

\section{Models and Methods\label{m_and_m}}

\subsection{Self-avoiding walk (SAW)}
We denote points on a $d$-dimensional cubic lattice by $\omega(i) \in \mathbb{Z}^d\ (i=0,1,2,\cdots)$ and a set of points by a path $\omega$. An $N$-step random walk (RW) is defined by $\omega=(\omega(0),\omega(1),\cdots,\omega(N))$ starting from the origin of the lattice with the constraint $|\omega(i+1)-\omega(i)|=1\ (i=0,1,\cdots,N-1)$. We denote the universal set of RWs as $\{\omega\}_\text{RW}$.
Since each site has $2d$ nearest neighbors, the total number of $N$-step RWs is exactly $(2d)^N$.
In the case of a SAW, a further constraint imposed by $\omega(i)\neq \omega(j)$ for all $i\neq j$ and we denote the universal set of SAWs as $\{\omega\}_\text{SAW}$.
This constraint makes it difficult to exactly count the total number of $N$-step SAWs $c_N$.

\subsection{Multicanonical Monte Carlo method}
Two enumeration methods proposed in this paper are based on the same sampling technique called the multicanonical Monte Carlo method~\cite{BergNeuhaus1991,BergNeuhaus1992}. If we define Hamiltonian $\mathcal{H}$ as a function which maps a path on a $d$-dimensional lattice to energy, $\mathcal{H}:\ \omega_a \to E_a$, we can introduce an energy structure into $\{\omega\}_\text{RW}$ and $\{\omega\}_\text{SAW}$.
Using the multicanonical method, we can estimate the number of states $\Omega(E)$ accurately over a wide range of energy, as we will explain in the following sections.

Introducing a weight $W(E)$ as a function of $E$ into the Markov chain Monte Carlo, we define the transition probability from a path $\omega_a$ to another path $\omega_b$
\begin{equation}
p(\omega_a \to \omega_b) = 
\min\left[
\frac{W(E_b)}{W(E_a)},1
\right],
\end{equation}
where $E_a$ and $E_b$ are energy of $\omega_a$ and $\omega_b$, respectively. 
To calculate thermodynamic values at specific inverse temperature $\beta$, we can use the Metropolis' method with $W(E)\propto \exp(-\beta E)$.
In the multicanonical method, the energy space of $W(E)$ is divided into $n$ bins and inverse temperature $\beta_i\ (i=1,2,\cdots ,n)$ is introduced to each bin.
$W(E)$ of the $i$th bin is set to be proportional to $\exp(-\beta_i E)$, and joint parameters $\alpha_i\ (i=1,2,\cdots,n-1)$ are introduced to connect $W(E)$ continuously at the boundaries of the bins.
By modifying $\beta_i\ (i=1,2,\cdots ,n)$ and $\alpha_i\ (i=1,2,\cdots ,n-1)$, we determine $W(E)$ to be approximately proportional to the inverse of the number of states $1/\Omega(E)$.
If the number of energy levels $n_E$ is finite and $n$ is equal to $n_E$, which is true throughout our study, the multicanonical method is identical to the entropic sampling~\cite{Lee1993}.

In our study, we built up $W(E)$ using the Wang-Landau method~\cite{WangLandau2001,WangLandau2001PRE}, the detailed procedure of which is given in the Appendix.
If we give a transition probability using this $W(E)$, the Markov chain Monte Carlo produces a flat histogram $H(E)$.
We can obtain $H(E)/W(E)\propto \Omega(E)$ with high accuracy over a wide range of the energy scale. In order to estimate the number of states, $\Omega(E)$, we need to introduce an absolute scale.
In Sec.~\ref{sec:go} and \ref{sec:dj}, we introduce two different absolute scales.

\subsection{Multi-self-overlap ensemble} \label{subsec:MSOE}
To efficiently sample SAWs from $\{\omega\}_\text{SAW}$, we can use the multi-self-overlap ensemble (MSOE)~\cite{IbaKikuchi1998,ChikenjiIba1999}, which is an extended version of the multicanonical method.
Relaxing the self-avoiding conditions, the MSOE makes it possible to explore the configuration space faster than the case that the self-avoiding conditions are strictly kept despite the extra configuration space.
The reason for the fast exploration is that transition paths from one SAW to another increase via configurations with one intersection or more.
Let $V$ be the number of the overlaps of a path.
If $k$ points of a path $ (k\ge 3) $ are on the same site, we define that there are $k-1$ overlaps on that site.
Note that, in the original paper, $V$ is defined as penalty for overlaps, and they used $(k-1)^2$ instead of $k-1$.
Using $V$ and the prescribed cutoff, $V_\text{cut}$, we denote the original configuration space and the expanded configuration space as $\{\omega\}^{V=0}_\text{RW}$ and $\{\omega\}^{V\le V_\text{cut}}_\text{RW}$, respectively.

In the same way as with the multicanonical method, we build up the weight function $W(E,V)$.
By using the transition probability
\begin{equation}
p(\omega_a \to \omega_b) = 
\min\left[
\frac{W(E_b,V_b)}{W(E_a,V_b)},1
\right],
\end{equation}
we obtain a two-dimensional flat histogram, $H(E,V)$.
The number of $\{\omega\}_\text{SAW}$ can be obtained from the equation $\Omega(E)\propto H(E,0)/W(E,0)$.
We used the MSOE in the first enumeration method using the G\={o} model.
The idea of using the number of overlaps as a variable is shared by the second enumeration method using the Domb-Joyce model, which has the fully expanded configuration space $\{\omega\}_\text{RW}$.

\section{Multicanonical simulation of the G\={o} model
\label{sec:go}}
\subsection{G\={o} model}
In the first enumeration method, we estimate $c_N$ by multicanonical simulations of the G\={o} model~\cite{TaketomiUedaGo1975,GoTaketomi1978,Go1983}, which was introduced originally to investigate the protein folding problem theoretically. 
The G\={o} model is defined as a SAW with specialized interactions that gives one conformation $\omega_\text{native}$, called the native structure, as a unique ground state apart from the trivial spatial symmetry.
Using two indices of points of $\omega$, $i$ and $j$, a native contact pair is defined by a pair of $i$ and $j\ (j>i+1)$ which satisfies the condition $|\omega_\text{native}(i)-\omega_\text{native}(j)|=1$.
The Hamiltonian of the intrachain interactions can be written as
\begin{equation}
\mathcal{H} (\omega)=\sum_{(i,j)\in\{\text{native contact pairs}\}}-\varepsilon\ \delta(|\omega(i)-\omega(j)|,1).
\end{equation}
If we define $n_\text{ncp}$ as the total number of native contact pairs, the ground state energy $E_\text{GS}$ is written as $-\varepsilon n_\text{ncp}$.

\subsection{Designing a native state}
We calculate $g(E)=H(E,0)/W(E,0)$ of the G\={o} model using the MSOE.
In order to estimate $\Omega(E)$ from $g(E)$, we need to know the number of states of at least one bin of $\Omega(E)$.
We design a $\omega_\text{native}$ that is suitable for that purpose, leaving aside the protein folding problem.

\begin{figure}[htb]
\begin{center}
\begin{minipage}{9pc}
\includegraphics[width=9pc]{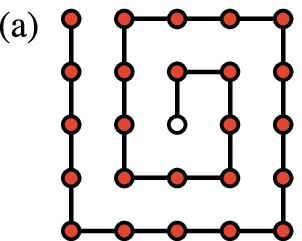}
\end{minipage}\hspace{7pc}
\begin{minipage}{9pc}
\includegraphics[width=9pc]{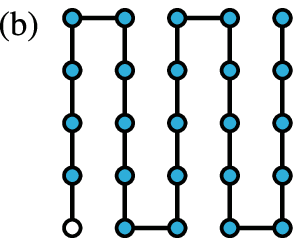}
\end{minipage} 
\caption{\label{fig:go}
Two types of native states of the G\={o} model ($N=24$): (a) roll and (b) beta shape. The start point, $\omega(0)$, of each chain is shown as a white circle.}
\end{center}
\end{figure}

\begin{figure}[htb]
\begin{center}
\begin{minipage}{15pc}
\includegraphics[width=16pc]{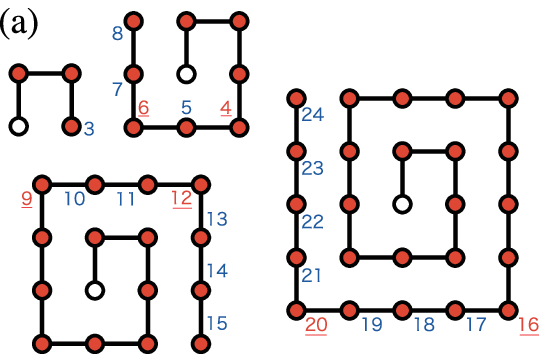}
\end{minipage}\hspace{4pc}%
\begin{minipage}{15pc}
\includegraphics[width=16pc]{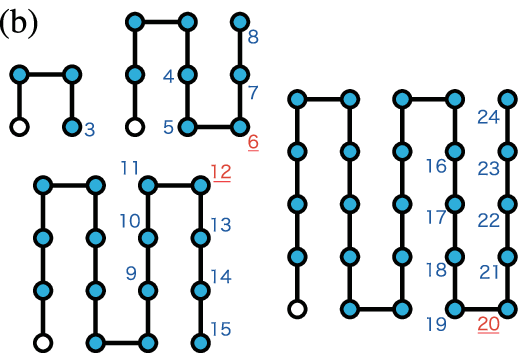}
\end{minipage} 
\caption{\label{fig:go_numbering} 
(a) Roll- and (b) beta-shaped native states for the G\={o} models with $N=3-24$.
The start point, $\omega(0)$, of each chain is shown as a white circle. The steps with blue numbers have 8 ground states and the steps with underlined red numbers have 16 ground states.}
\end{center}
\end{figure}

Selecting a compact structure which can maximize the number of the native contact pairs for $\omega_\text{native}$, we can easily count the number of ground states.
In the case of $d=2$ and $N=24$, for example, if we choose the configuration shown in figure~\ref{fig:go}-(a) ("roll" shape) or (b) ("beta" shape) as a native structure, the number of ground states is 8 for both cases considering spatial symmetry.
Roll-shaped or beta-shaped native states of any $N$ are made from the following rules.
\begin{enumerate}
\item[I.] Find $L\ (\in \mathbb{N})$ which satisfies the condition $\sqrt{N}<L\le \sqrt{N}+1$.
\item[II.] Draw the $L \times L$ compact structure using a roll or beta shape (figure \ref{fig:go_numbering} for $L=2,3,4,5$).
\item[III.] Select $N$ steps from the start point $\omega(0)$. This can be used as $\omega_\text{native}$ for the $N$-step G\={o} model.
\item[IV.] The number of ground states are determined as below.
\end{enumerate}
In the case of a roll shape,
\begin{equation}
\Omega(E_\text{GS})=
\begin{cases}
16 & \mbox{when } N=(\big\lfloor \sqrt{N} \big\rfloor)(\big\lceil \sqrt{N} \big\rceil)\\
8 & \mbox{otherwise},
\end{cases}\notag
\end{equation}
and in the case of a beta shape,
\begin{equation}
\Omega(E_\text{GS})=
\begin{cases}
16 & \mbox{when } N=L^2\\
8 & \mbox{otherwise},
\end{cases}\notag
\end{equation}
where $\lfloor\ \rfloor$ is the floor function and $\lceil\ \rceil$ is the ceiling function.
In the cases of $\Omega(E_\text{GS})=16$, there is an extra double degeneracy at the end of the paths.
The roll-shaped and beta-shaped native states for $N\le 24$ are illustrated in figure~\ref{fig:go_numbering}-(a) and (b), respectively.

\subsection{Scale fixing by the number of ground states}
\begin{figure}[h]
\begin{center}
\begin{minipage}{15pc}
\includegraphics[width=15pc]{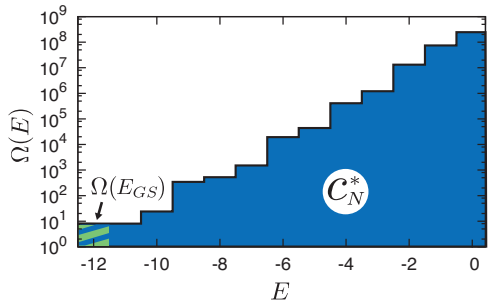}
\end{minipage}
\hspace{2pc}%
\begin{minipage}{14pc}
\caption{\label{fig:scale_fixing_go}
The idea of scale fixing demonstrated with a roll-shaped $\omega_\text{native}$ of the G\={o} model ($N=19$). This model includes $\{\omega\}_\text{SAW}$ as the whole of its configuration space and also includes a scale fixing set $\{\omega\}_\text{GS}$ as a part.}
\end{minipage}
\end{center}
\end{figure}

If we design $\omega_\text{native}$ with specified $\Omega(E_\text{GS})$, like a roll or beta shape, the estimated number of states, $\Omega^*(E)$ (we denote estimated values calculated by simulations with asterisk ($*$)), can be calculated from $g(E)$ as 
\begin{equation}
\Omega^*(E)=\frac{\Omega(E_\text{GS})}{g(E_\text{GS})} g(E) = \lambda_\text{G\={o}}\ g(E).
\end{equation}
Here we defined the scale fixing factor, $\lambda_\text{G\={o}}$, by $\Omega(E_\text{GS})/g(E_\text{GS})$.
Since the $N$-step G\={o} model has the same configuration space as $\{\omega\}_\text{SAW}$, the whole sum of $\Omega(E)$ is equal to $c_N$, and the estimated number of $N$-step SAWs is given by
\begin{equation}
c^*_N=\sum_E \Omega^*(E).
\end{equation}
The scale of $\Omega(E)$ is determined by the set of the ground states $\{\omega\}_\text{GS}$, which we call the scale fixing set.
The idea of the scale fixing with the G\={o} model is illustrated in figure~\ref{fig:scale_fixing_go}.
This model includes $\{\omega\}_\text{SAW}$ as the whole of its configuration space and, if it has a well-defined native state, it also includes a scale fixing set $\{\omega\}_\text{GS}$ as a part.

\section{Multicanonical simulation of the Domb-Joyce model
\label{sec:dj}}
\subsection{Domb-Joyce model}
In the second enumeration method, we estimate $c_N$ by multicanonical simulations of the Domb-Joyce model~\cite{DombJoyce1972}.
The configuration space of the Domb-Joyce model is $\{\omega\}_\text{RW}$.
In the Domb-Joyce model, each path is weighted with the factor
\begin{equation}
\prod^{N-2}_{i=0}\prod^N_{j=i+2} (1-u\delta_{ij}),
\end{equation}
where $u$ is a parameter of the model. If we replace $u$ by $1-\exp(-\beta J)$, each weight is identical to the Boltzmann factor of the Hamiltonian, which is written as
\begin{equation}
\mathcal{H} (\omega)=\sum_{i<j} J\ \delta(\omega(i),\omega(j))=JV_\mathrm{DJ},
\end{equation}
where $V_\mathrm{DJ}$ is the number of intersections of a path.
In the discussion below, we assume $J>0$, which means a pair of crossed steps as a repulsive interaction of the strength $J$.
In the limit of $\beta\to 0\ (w\to 0)$, all the configurations are equally weighted and the Domb-Joyce model corresponds to $N$-step RWs.
In the opposite limit, $\beta \to \infty\ (w\to 1)$, the Domb-Joyce model reduces to $N$-step SAWs.

\subsection{Modified Domb-Joyce model} \label{subsec:modifiedDJ}
In the actual simulation, we used a slightly modified model of the Domb-Joyce model in order to narrow the range of energy. 
In the modified model, we use $V$ in section~\ref{subsec:MSOE} instead of $V_\mathrm{DJ}$.
We explain their difference.
If we introduce a number of intersections of each site on a lattice as $V (x, y) $, we calculate $V_\mathrm{DJ}$ or $V$ by
\begin{equation}
\sum_{x=-N}^{N}\sum_{y=-N}^{N} V(x,y).
\end{equation}
where $V (x, y) =0$ at sites without points of a path.
In the original Domb-Joyce model, if there are $k$ points $(k\ge 2) $ on the same site $ (x_s,y_s) $, $V(x_s,y_s)$ is given by $V(x_s,y_s)=k(k-1)/2$.
We, however, use the definition given by $V(x_s,y_s)=k-1$.
This modification does not change the properties in $\beta \to 0$ and $\beta \to \infty$ of the Domb-Joyce model.

\subsection{Scale fixing by the total number of $N$-step random walks}
\begin{figure}[h]
\begin{center}
\begin{minipage}{15pc}
\includegraphics[width=15pc]{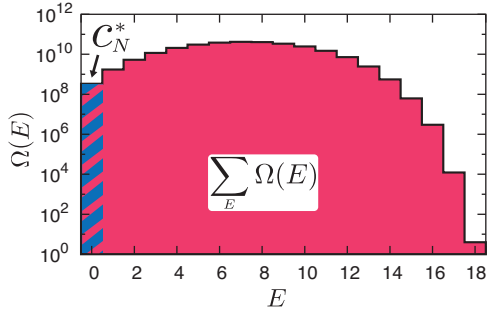}
\end{minipage}
\hspace{2pc}%
\begin{minipage}{14pc}
\caption{\label{fig:scale_fixing_dj}
The idea of scale fixing demonstrated with the modified Domb-Joyce model ($N=19$).
This model includes $\{\omega\}_\text{SAW}$ as a part of its configuration space and also includes a scale fixing set $\{\omega\}_\text{RW}$ as a whole.}
\end{minipage}
\end{center}
\end{figure}

From the multicanonical simulations of the modified Domb-Joyce model, we can obtain $g(E)=H(E)/W(E)\ (E=JV)$ of the conformation space $\{\omega\}_\text{RW}$. Since the total number of $\{\omega\}_\text{RW}$ is exactly known as $(2d)^N$ in the $d$-dimensional cubic lattice, the universal set of the modified Domb-Joyce model $\{\omega\}_\text{RW}$ can be used as a scale fixing set.
Using a scale fixing factor $\lambda_\text{DJ}=\sum_E \Omega(E)/\sum_E g(E)$, we can calculate the estimated number of states $\Omega^*(E)$ as
\begin{equation}
\Omega^*(E)=\lambda_\text{DJ}\ g(E).
\end{equation}
Then, $c^*_N$ is obtained by
\begin{equation}
c^*_N=\Omega^*(0).
\end{equation}
The idea of the scale fixing with the modified Domb-Joyce model is illustrated in figure~\ref{fig:scale_fixing_dj}. 
In contrast with the G\={o} model, the modified Domb-Joyce model includes $\{\omega\}_\text{SAW}$ as a part of its configuration space and also includes a scale fixing set $\{\omega\}_\text{RW}$ as a whole.

\section{Results and Discussions\label{results}}
\begin{figure}[h]
\begin{center}
\begin{minipage}{15pc}
\includegraphics[width=15pc]{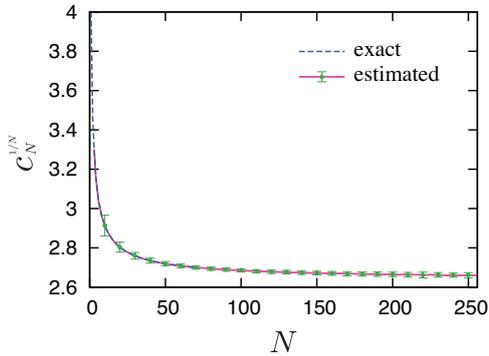}
\end{minipage}
\hspace{2pc}%
\begin{minipage}{15pc}
\caption{\label{fig:number}
A plot of $(c^*_N)^{1/N}$ (pink) and $c^{1/N}_N$ (blue).
The error bars were shown every 10 steps and their values were magnified by a factor of $20$.}
\end{minipage}
\end{center}
\end{figure}

We estimated $c_N$ up to 143 steps with the G\={o} model and up to 256 steps with the modified Domb-Joyce model with the same amount of computational effort.
Thus, we conclude that the modified Domb-Joyce model is more efficient than the G\={o} model for enumerating SAWs.
The results are shown in table~\ref{table1} along with the known exact numbers~\cite{Jensen2004} up to 71 steps.
$c^*_N$ agree with the exact numbers within the statistical error.
It should be noted that the present methods are statistically unbiased.
These enumeration methods are highly efficient and successfully counted large numbers up to $10^{108}$.
\begin{table}
\caption{A table of the estimated number of $N$-step SAWs $c^*_N$ with the $N$-step exact number of SAWs $c_N$. 
The exact and estimated numbers are rounded off to 5 digits or less.
The numbers in parentheses are the statistical errors for the last digits of the estimated numbers.
\label{table1}}
{\footnotesize
\begin{center}
\begin{tabular}{rrrrrrr}
\hline\hline
$N$ & $L$ & \multicolumn{1}{c}{$c_N$} & \multicolumn{4}{c}{$c^*_N$}\\
& & \multicolumn{1}{c}{Exact} & & \multicolumn{1}{c}{G\={o} (roll)} &\multicolumn{1}{c}{G\={o} (beta)} & \multicolumn{1}{c}{Domb-Joyce}\\
\hline 3 & 2 & $36$ & & $3.602(1)\times 10^{1}$ & $3.60(2)\times 10^{1}$ & $3.60(3)\times 10^{1}$\\
8 & 3 & $5916$ & & $5.92(1)\times 10^{3}$ & $5.92(3)\times 10^{3}$ & $5.92(5)\times 10^{3}$\\
15 & 4 & $6.4166\times 10^{6}$ & & $6.42(3)\times 10^{6}$ & $6.44(4)\times 10^{6}$ & $6.42(6)\times 10^{6}$\\
24 & 5 & $4.6146\times 10^{10}$ & & $4.62(3)\times 10^{10}$ & $4.62(2)\times 10^{10}$ & $4.61(4)\times 10^{10}$\\
35 & 6 & $2.2525\times 10^{15}$ & & $2.24(2)\times 10^{15}$ & $2.26(1)\times 10^{15}$ & $2.25(2)\times 10^{15}$\\
48 & 7 & $7.5014\times 10^{20}$ & & $7.48(8)\times 10^{20}$ & $7.55(7)\times 10^{20}$ & $7.51(7)\times 10^{20}$\\
63 & 8 & $1.7155\times 10^{27}$ & & $1.72(3)\times 10^{27}$ & $1.72(3)\times 10^{27}$ & $1.74(2)\times 10^{27}$\\
71 & 8 & $4.1909\times 10^{30}$ & & $4.3(2)\times 10^{30}$ & $4.24(8)\times 10^{30}$ & $4.20(5)\times 10^{30}$\\
80 & 9 &  & & $2.74(8)\times 10^{34}$ & $2.71(6)\times 10^{34}$ & $2.68(3)\times 10^{34}$\\
99 & 10 &  & & $2.9(3)\times 10^{42}$ & $3.0(1)\times 10^{42}$ & $2.91(4)\times 10^{42}$\\
120 & 11 &  & & $2.5(2)\times 10^{51}$ & $2.16(9)\times 10^{51}$ & $2.17(4)\times 10^{51}$\\
143 & 12 &  & & $1.4(4)\times 10^{61}$ & $1.19(8)\times 10^{61}$ & $1.12(3)\times 10^{61}$\\
168 & 13 &  & &  &  & $4.1(1)\times 10^{71}$\\
195 & 14 &  & &  &  & $9.9(4)\times 10^{82}$\\
224 & 15 &  & &  &  & $1.77(8)\times 10^{95}$\\
255 & 16 &  & &  &  & $2.1(1)\times 10^{108}$\\
256 & 16 &  & &  &  & $6.2(4)\times 10^{108}$\\
\hline\hline
\end{tabular}
\end{center}
}
\end{table}
$c^{1/N}_N$ is shown in figure~\ref{fig:number}.

\begin{figure}[htb]
\begin{center}
\begin{minipage}{23pc}
\includegraphics[width=24pc]{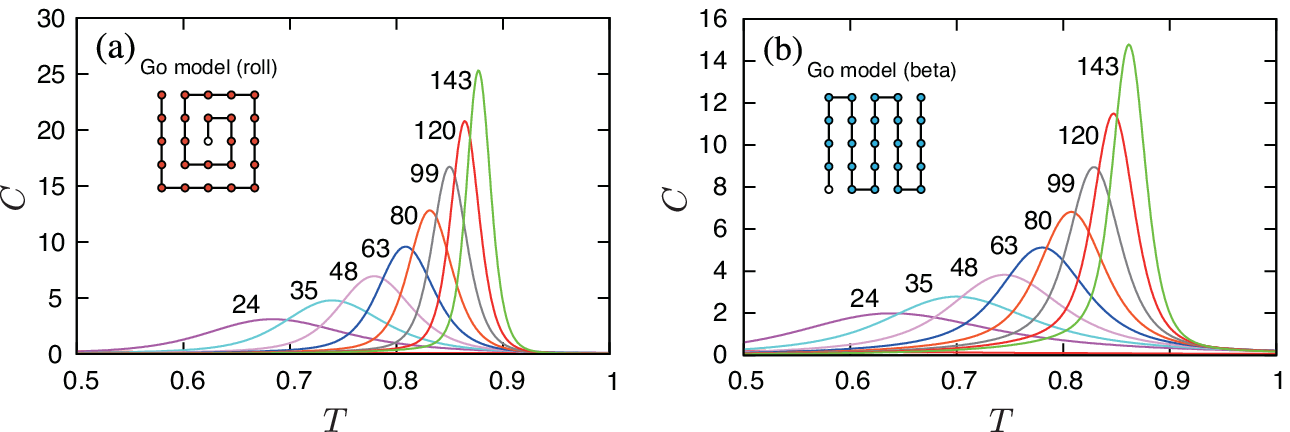}
\caption{\label{fig:C_go}
Temperature dependence of the specific heats of the G\={o} model with (a) roll- and (b) beta-shaped native states.}
\end{minipage}\hspace{1.5pc}%
\begin{minipage}{12pc}
\includegraphics[width=11pc]{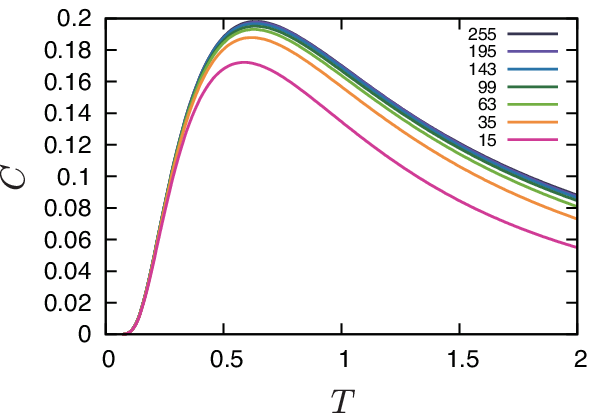}
\caption{\label{fig:C_DJ}
Temperature dependence of the specific heats of the modified Domb-Joyce model.}
\end{minipage} 
\end{center}
\end{figure}

Using $g(E)$ calculated from the multicanonical simulations, we calculated specific heats of the two models (see figure~\ref{fig:C_go} and \ref{fig:C_DJ}).
Based on the size dependence of the specific heats of the G\={o} model, 
we expect that the model exhibits a first-order phase transition at the long chain limit.
We consider that this transition can partly explain the reason for the performance difference between the two developed methods, and we suggest that the performance of the statistical enumeration can be evaluated by the thermodynamic behavior of the models.

\section*{Acknowledgments}
This work was supported by the Global COE Program Core Research and Engineering of Advanced Materials-Interdisciplinary Education Center for Materials Science, MEXT, Japan.

\newpage

\section*{Appendix. Algorithm of the developed enumeration methods for $N$-step SAWs}
A pseudo-code of the Wang-Landau method and the multicanonical method with a flow chart of the whole procedure to calculate $c^*_N$ with statistical errors.
\vspace{-1.5mm}
\begin{multicols}{2}
\hrule height 0.08pc
\vspace{1mm}
{\footnotesize
\noindent {\bf Algorithm 1.} The Wang-Landau method and the multicanonical method (lines with \textcolor{red}{WL} were skipped in the latter algorithm) %
\vspace{1mm}
\hrule 
\vspace{1mm}
\noindent {\bf Initialization}
\vspace{1mm}
\hrule 
\begin{center}
\begin{algorithmic}[1]
\For{the whole range of $E$}
	\State $W(E) \leftarrow 1$
	\State $H(E) \leftarrow 0$
\EndFor
\State make the first conformation $\omega_0$ and calculate its energy $E_0$
\State $f\leftarrow e$ \Comment $f$ is a modification factor of the weight function $W(E)$.
\State $T\leftarrow 10^7$ \Comment $T$ is a cycle of judgement for updating $f$.\State $\lambda_\text{flat}\leftarrow 0.7\sim 0.95$
\State \Comment $\lambda_\text{flat}$ is a flatness parameter of judgment for updating $f$.
\State \Comment These values of the initial settings are typical values.
\algstore{break1}
\end{algorithmic}
\end{center}
\vspace{1mm}
\hrule 
\vspace{1mm}
\noindent {\bf Functions}
\vspace{1mm}
\hrule 
\vspace{-1mm}
\begin{center}
\begin{algorithmic}[1]
\algrestore{break1}
\Function{Move}{$\omega$}
	\State make a candidate state $\omega^\prime$ from $\omega$
	\State returning $\omega^\prime$
\EndFunction
\Statex
\Function{Energy}{$\omega$}
	\State calculate energy $E$ of $\omega$ from $\mathcal{H}$
	\State returning $E$
\EndFunction
\Statex
\Function{Judge}{$E_a$,$E_b$,$W(E)$}
	\State generate a uniform random number $r\in [0,1)\ (r\in\mathbb{R})$
	\If{$r<\frac{W(E_b)}{W(E_a)}$}
		\State returning $1$
	\Else
		\State returning $0$
	\EndIf
\EndFunction
\Statex
\Function{Update}{$H(E)$}
	\State $U \leftarrow 1$
	\State calculate the average of the histogram $H(E)$, $\overline{H}$
	\For{the whole range of $E$}
	\State \Comment In some cases, the range of $E$ is limited by hand
	\If{$H(E)<\overline{H}\cdot \lambda_\text{flat}$}
		\State $U \leftarrow 0$
	\EndIf
	\EndFor
\EndFunction
\Statex
\Function{Initialize}{$H(E)$}
	\For{the whole range of $E$}
		\State $H(E) \leftarrow 0$
	\EndFor
\EndFunction
\algstore{break2}
\end{algorithmic}
\end{center}
\vspace{1mm}
\hrule 
\vspace{1mm}
\noindent {\bf Main part}
\vspace{1mm}
\hrule 
\begin{center}
\begin{algorithmic}[1]
\algrestore{break2}
\While{$f-1>10^{-8}$}
\For{$t:=0$ to $T-1$}
	\State $\omega^\prime_t \leftarrow$ \Call{Move}{$\omega_t$}
	\State $E^\prime_t \leftarrow$ \Call{Energy}{$\omega^\prime_t$}
	\State $J \leftarrow$ \Call{Judge}{$E_t$,$E^\prime_t$,$W(E)$}
\If{$J=1$}
	\State $\omega_{t+1} \leftarrow \omega^\prime_t$
	\State $E_{t+1} \leftarrow E^\prime_t$
\Else 
	\State $\omega_{t+1} \leftarrow \omega_t$
	\State $E_{t+1} \leftarrow E_t$
\EndIf
	\State $W(E_{t+1}) \leftarrow W(E_{t+1})/f$
	\Comment{\textcolor{red}{WL}}
	\State $H(E_{t+1}) \leftarrow H(E_{t+1})+1$
\EndFor
	\State $U \leftarrow $ \Call{Update}{$H(E)$}
	\Comment{\textcolor{red}{WL}}
\If{$U=1$} \Comment{\textcolor{red}{WL}}
	\State $f \leftarrow f^{\frac{1}{2}}$
	\Comment{\textcolor{red}{WL}}
\EndIf
\Comment{\textcolor{red}{WL}}
	\State \Call{Initialize}{$H(E)$}
	\Comment{\textcolor{red}{WL}}
\EndWhile
\end{algorithmic}
\end{center}
}
\vspace{-1mm}
\hrule height 0.08pc
\begin{figurehere}
\begin{center}
\includegraphics[width=9pc]{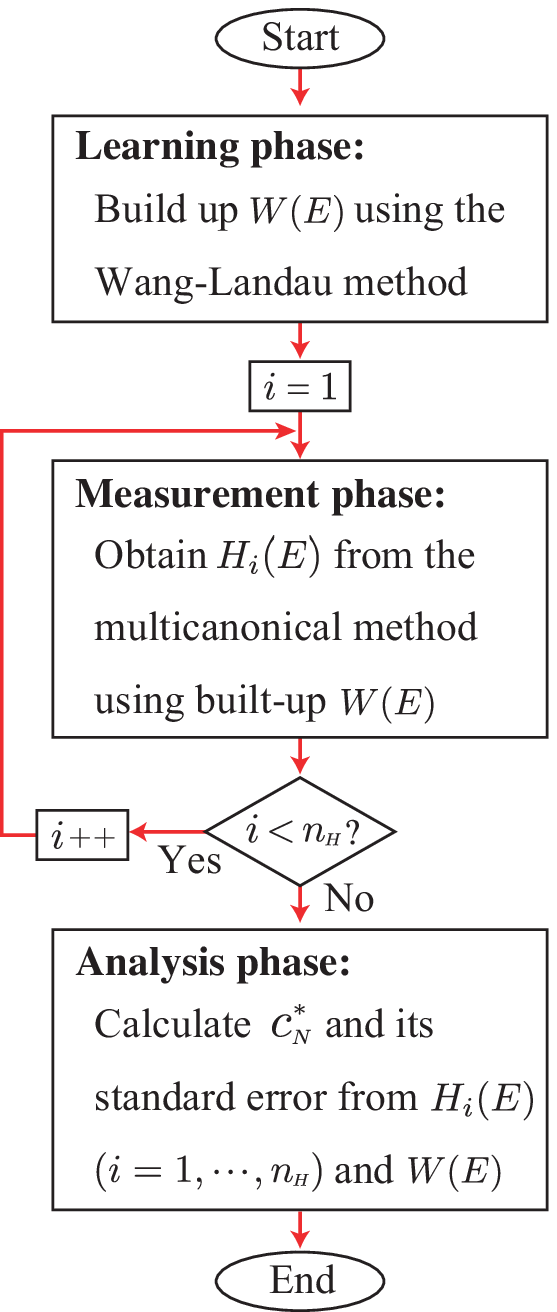}
{\footnotesize
\caption{Flow chart of the developed enumeration methods for $N$-step SAWs. $n_H$ is the number of Histograms.}}
\label{fig:flowchart}
\end{center}
\end{figurehere}
\end{multicols}

\end{document}